\documentclass[12pt,a4paper]{article}

\usepackage[left=1in,right=1in,top=1in,bottom=1in]{geometry}
\usepackage{setspace}
\usepackage[hidelinks]{hyperref}

\usepackage{amsmath, amssymb}
\usepackage{graphicx}
\usepackage{hyperref}
\usepackage{booktabs}
\usepackage{enumitem}
\usepackage{caption}
\usepackage{subcaption}
\usepackage{algorithm}
\usepackage[noend]{algorithmic}
\usepackage{multirow}
\usepackage{wrapfig}
\usepackage{natbib}
\usepackage{authblk}

\begin{document}

\begin{center}
    {\Large\bfseries Optimizing Sponsored Humanitarian Parole}\\
    \vspace{1em}  
\end{center}
\noindent
\begin{center}
\begin{tabular}[t]{c}
    Fatemeh Farajzadeh \\
    Data Science, Worcester Polytechnic Institute \\
    \vspace{0.5em} \\ 
    
    Ryan B. Killea \\
    Data Science, Worcester Polytechnic Institute \\
    \vspace{0.5em}  \\ 
    
    Alexander Teytelboym\\
    Department of Economics, University of Oxford \\
    \vspace{0.5em} \\ 
    
    Andrew C. Trapp \\
    WPI Business School, Worcester Polytechnic Institute \\
    Data Science, Worcester Polytechnic Institute 
\end{tabular}
\end{center}
\vspace{2em}

\begin{abstract}
    The United States has introduced a special humanitarian parole process for Ukrainian citizens in response to Russia's 2022 invasion of Ukraine. To qualify for parole, Ukrainian applicants must have a sponsor in the United States. 
    In collaboration with HIAS, a refugee resettlement agency involved in the parole process, we deployed RUTH (Refugees Uniting Through HIAS), a novel algorithmic matching system that is driven by the relocation preferences of refugees and the priorities of US sponsors.
    RUTH adapts \cite{Thakral} Multiple-Waitlist Procedure (MWP) that combines the main First-In/First-Out (FIFO) queue with location-specific FIFO queues in order to effectively manage the preferences of refugees and the supply of community sponsors.     
    In addition to refugee preferences and sponsor priorities, RUTH incorporates various feasibility considerations such as community capacity, religious, and medical needs. The adapted mechanism is envy-free, efficient and strategy-proof for refugees.
    Our analysis reveals that refugee preferences over locations are diverse, even controlling for observables, by demonstrating the difficulty of solving a much simpler problem than modeling preferences directly from observables.
    We use our data for two counterfactual simulations. First, we consider the effects of increased waiting times for refugees on the quality of their matches. We find that with a periodic Top Trading Cycles algorithm, increasing period length from 24 days to 80 days, improves average rank of a refugee's match from 3.20 to 2.44. On the other hand, using the available preference data RUTH achieved an average rank of 4.07 with a waiting time of 20 days.
    Second, we estimate the arrival rates of sponsors in each location that would be consistent with a long-run steady state. We find that more desirable locations (in terms of refugee preferences) require the highest arrival rates suggesting that preferences might be a useful indicator for investments in sponsorship capacity.
    Our study highlights the potential for preference-based algorithms such as RUTH to improve the efficiency and fairness of other rapidly-deployed humanitarian parole processes.
\end{abstract}

\section{Introduction}

Around a quarter of the population in Ukraine was displaced by the 2022 Russian invasion, creating the largest refugee crisis in Europe since World War II~\cite{UNHCR1}: more than 8 million Ukrainians have crossed international borders~\cite{UNHCR2}.  Since the start of the war, hundreds of thousands of Ukrainians have been granted temporary asylum in the United States (US). The US government set up the Uniting for Ukraine (U4U) program to provide a pathway for Ukrainian citizens to temporarily remain in the US for up to two years under federal humanitarian parole status.\footnote{Some Ukrainian applicants apply directly from Ukraine and are therefore technically considered ``internally displaced'' rather than ``refugees'' (as refugees must cross an international border). Nevertheless, we will refer to all Ukrainian applicants as ``refugees'' fleeing a war.} However, prior to applying for parole, Ukrainian refugees are required to have private sponsors who commit to providing them with financial, social, and residential support. Many organizations, including NGOs tasked with conventional refugee resettlement, have been helping Ukrainian refugees find sponsors. Over 100,000 Ukrainians have already successfully applied for parole via U4U~\cite{Over100}.

Ample empirical evidence reveals that the initial placement of refugees significantly impacts their long-term welfare outcomes~\cite{aaslund2007and,aaslund2010important,damm2014neighborhood}. Several studies have attempted to improve refugee outcomes by matching refugees to locations where they are most likely to quickly find employment using rich observable information about refugees ~\cite{ban18,bansak2022dynamic,ahani21,ahani23}. Such outcome-based systems, however, incorporate neither the preferences of refugees nor the priorities of communities that host them (e.g., ~\cite{jon16,delacretaz2019matching}). If indeed there were substantial diversity in refugee preferences over communities (for example, some refugees may require access to higher education, while others might prioritize social connections), then preference elicitation might be socially valuable as it could help to improve the relocation process using refugees’ private information. 

In this paper, we describe RUTH (Refugees Uniting Through HIAS), a novel algorithmically driven platform that matches refugees to communities by explicitly considering the preferences of eligible Ukrainian refugees with the priorities of US sponsors. The matching mechanism in RUTH adapts and implements the Multiple-Waitlist Procedure (MWP) due to~\citet{Thakral}. In partnership with HIAS, a US-based refugee agency, we incorporate staff-informed features that address fairness factors relevant to relocating refugees while maintaining strategy-proofness, efficiency, and elimination of justified envy properties of the original procedure. In particular, we adapt the MWP by including 
additional feasibility constraints pertaining to sponsor hosting capacity, religious compatibility, and medical needs. We maintain that the MWP achieved all of the key objectives HIAS set before us in this project, including rapid matching and giving refugees agency and choice.

The matching algorithm operates as follows. Sponsors enter the algorithm through the locations with which they are associated, while refugee families enter the algorithm through the main FIFO queue. There is also a separate FIFO queue for each location. The first available sponsor is matched to the highest-priority feasible family in the locational queue associated with the sponsor. However, if the locational queue is empty or there is no feasible match, the available sponsor is offered to the highest-priority feasible family in the main queue. Such a family can either immediately accept the offer or opt to join the FIFO queue of their preferred location based on estimated waiting times provided by the agency. Once in the locational queue, the family simply waits its turn to be matched to a feasible sponsor.

RUTH is implemented and deployed in Eastern Europe and Washington DC via a custom-built, secure web app.
As of April 2023, RUTH has helped to successfully relocate over 40 families  (136 persons) to major metropolitan cities in the states of New York, New Jersey, California, and Colorado. Successfully relocated families on average waited just over two weeks to finalize a match.
We find that refugee preferences in our sample over locations are diverse, mainly favoring large metropolitan areas.  
This highlights the value of eliciting truthful preferences in preference-based matching, which can improve relocation by leveraging private information.

We conduct two counterfactuals to better understand the performance of RUTH. In the first, we evaluate the quality of matches and average refugee waiting times that RUTH produces by comparing them with periodic Top Trading Cycles (TTC) batched at varying frequencies. We observe that when we reduce waiting time for families (by increasing the frequency of periodic TTC), match quality decreases because there are fewer sponsors from desirable locations available to match. However, we find that the MWP in RUTH substantially outperforms periodic TTC in terms of waiting time, given its level of match quality.
The second counterfactual aims to determine long-run steady arrival rates of sponsors by simulating the refugee arrival and matching process using the observed preference distribution. We find that by tuning sponsor arrivals across locations according to the preferences of refugees, we can drastically improve long-run matching rates. This insight might help resettlement agencies allocate investment for sponsor recruitment more efficiently.

The rest of this paper is organized as follows. Section 2 provides background on related literature. Section 3 describes the model, the MWP, and the properties of the MWP. Section 4 describes the details of our implementation of the MWP in RUTH. Section 5 offers a few immediate insights from our data about relocation destinations, refugee preferences and waiting times. Section 6 describes two counterfactual analyses. Section 7 concludes.

\section{Background and Literature Review}

The US follows a conventional, top-down approach for refugee resettlement, where the federal government partners in a public-private alliance with nine voluntary agencies, including HIAS, under the US Refugee Admissions Program (USRAP) \cite{wraps2023}.
Every year a quota on refugee admissions is set by the executive office, in consultation with Congress~\cite{StateRefugeeAdmissions}.
The USRAP oversees the process of admitting refugees recommended by the United Nations High Commissioner for Refugees (UNHCR) who undergo a thorough vetting process before being welcomed into the United States.
Under the Trump administration, the USRAP experienced considerable changes that led to a marked reduction in both the annual quota and number of admitted refugees~\cite{MPITrump}.
Meanwhile, as global refugee numbers continue to grow including those from the recent Ukrainian crisis, alternative pathways such as humanitarian parole are emerging to address the growing need for resettlement~\cite{USCISHP}.
Despite limitations such as the lack of long-term security, humanitarian parole offers certain advantages over traditional top-down resettlement, such as more expedient vetting, and the ability for refugees to express preferences over desired locations.

Present approaches to matching refugees to communities involve either manual methods or the use of machine learning to estimate employment probabilities and optimization to maximize outcomes~\cite{ahani21,ahani23,ban18,bansak2022dynamic,freund2023group}. However, these approaches ignore the preferences of refugees and the priorities of hosts. Several papers have suggested models for refugee matching with preferences in both static~\cite{delacretaz2019matching,jones2018local,andersson2020assigning,acharya2022combining,Caspari} and even dynamic settings~\cite{TommyAnderssonD}. However, until now, there has been no practical implementation of preference-based refugee matching systems.

Our algorithm is based on a contribution by \citet{Thakral}, who developed a dynamic matching algorithm in the context of social housing. Miraculously, his algorithm appears to have all the features that were desirable in our context and we adopted it with minor modifications. There are, of course, many dynamic matching models with features that are highly relevant in our setting \cite{Jacob22, doval2022dynamically, BlochCantala, ArnostiShi}.

\section{Model and the Matching Algorithm}

In this section, we provide an informal description of the model and the algorithm properties from \citet{Thakral}.\footnote{One can map Thakral's model to ours as follows. Our families are his ``applicants''; our locations are his ``buildings''; our sponsors are his ``units''. The first main modification is that we introduce exogenous, binary feasibility constraints between ``applicants'' and ``units''. From the applicants' (resp. buildings') perspective, this is equivalent to specifying preferences (resp. priorities) over buildings (resp. applicants) \emph{and an outside option of being unmatched}. These modifications do not, however, affect any of the properties of the matching algorithm described in Thakral's Proposition~2.}

There is a set $F$ of refugee families, a set $L$ of locations, and a set $C$ of sponsors. There is a strict, exogenous priority order over families\footnote{Such as time of joining the system; in practice, we experimented with other options, as shown in Appendix B.}. A family-location pair is either feasible, in which case it can be a possible match, or infeasible when no match is possible. Time is discrete. For each time period $t$, there is a history that specifies which sponsors have arrived previously and in that time period. In each time period, a set of sponsors arrives according to a time- and history-dependent probability distribution over the power set of remaining sponsors. We can assume that exactly one sponsor arrives in each time period.
A period-$t$ assignment is a function that matches each family to a sponsor or leaves the family (temporarily) unmatched in that period; an assignment is a collection of per-period assignments.

Families have strict and time-invariant preferences over pairs of locations and \emph{expected} waiting times.\footnote{We therefore assume that families are indifferent between feasible sponsors in a given location. One can extend Thakral's model to the setting in which sponsors in a given location are heterogeneous and uncertain from the point of view of families. See Appendix C} Longer expected waiting times for any sponsor adversely affect families. Families are assumed to be risk-neutral. As we will see, the key decision for a family is whether to accept a sponsor offer now versus being matched to a sponsor in a given location at a future date. A family's preferences can be summarized as a list of waiting times for each location where the difference between entry for location $\ell$ and entry for (less desirable) location $\ell'$ expresses the maximum extra time a family would wait for $\ell$ over $\ell'$.

Several modeling features are worth emphasizing. First, every match is irrevocable. Once the match is finalized, the family and the sponsor must undertake the time-consuming process of applying for the I-134 affidavit of support that demonstrates the financial responsibility of the sponsor, thus it would be unreasonable to create any uncertainty about the match. Second, while in principle refugees have preferences over pairs of sponsors and waiting times, they have almost no information about potential sponsors when they join the system.\footnote{However, as we will explain, refugees will be asked to decide whether to accept a match with a known sponsor or wait for an uncertain sponsor in another location.} Third, we assume that preferences are strict. In practice, even when refugees were provided ample information about locations, they were challenged to effectively prioritize their preferred choices over the alternative options.

A mechanism is a procedure that uses reported preferences, the exogenous priority orderings, and the history to choose an assignment in each period $t$. A mechanism is strategy-proof if the deviation from truthful preference revelation is not profitable along any possible arrival history.
Standard properties of efficiency and elimination of justified envy extend naturally to this setting. The ex-ante versions of these properties consider information available to families at some time period $t$. Let us call a pair of a location and expected waiting time an \emph{uncertain option}. A mechanism eliminates justified envy if whenever a family $f$ prefers uncertain option $o'$ of family $f'$ to its option $o$, it must be because $f'$ has a higher priority than $f$. A mechanism is efficient if whenever a family $f$ is rematched to an uncertain option $o'$ that is preferred to its current uncertain option $o$, it makes another family worse off.
The ex-post versions of efficiency and elimination of justified envy are defined with respect to realized arrivals of sponsors. In the definition, we can simply replace uncertain options with \emph{certain options}, that is, pairs of locations and certain waiting times. However, no mechanism is ex-post efficient and ex-post free of justified envy.

The Multiple Waitlist Procedure works as follows. There are $|L|+1$ FIFO queues: one queue $Q_\ell$ for each location $\ell\in L$ and one \emph{main} queue $Q$. Initially, refugee families join the main queue. Suppose in time period $t$, a sponsor $c$ arrives in location $\ell$. If the set $S$ of families in $Q_\ell$ for which $c$ is feasible is non-empty, then we match $c$ to the highest priority family in $S$. If $S$ is empty, then we offer $c$ to the highest-priority feasible family $f$ in $Q$ (if there is no such family, the sponsor $c$ waits until an arrival in $Q_\ell$). Family $f$ can either accept the offer, in which case we immediately match $c$ to $f$ or reject the offer and join its most preferred locational queue. In the latter case, we continue offering $c$ to next highest-priority feasible family. The pseudocode of the MWP implementation in RUTH is given in Algorithm~\ref{alg:mwp}.
\subsection{Properties of MWP versus HIAS objectives}\label{sec:vsobj}
The objectives of HIAS were: (1) to match refugees to the best possible sponsors; (2) to match refugees and sponsors as quickly as possible; (3) to respect the priority order of the families in the system; (4) to give refugees agency and choice; and (5) to build a transparent, simple, scalable and safe-to-use system for refugees and sponsors. While there is a tradeoff between objectives (1) and (2), as we explain below, our algorithm essentially achieved all of the objectives. MWP is efficient, which addresses Objective (1); eliminates justified envy, which addresses Objective (3), and is strategy-proof and requires no computational cost beyond feasibility checking because the matching process is serial, which addresses Objective (5).\footnote{The result would remain intact with heterogeneous sponsors within a given location subject to appropriately modifying envy-free and efficiency properties.
The intuition for the properties is the following. The mechanism is ex-ante Pareto efficient and strategy-proof because when the refugee is offered a sponsor in the main queue they are presented with a choice of a sponsor or their favorite locational queue and they cannot gain by selecting an inferior option given their preferences. As refugees are approached in priority order, justified envy (subject to feasibility constraints) is eliminated.} One can view MWP as an elegant variation of the serial dictatorship algorithm. In a serial dictatorship, an arriving sponsor would be immediately matched to the highest-priority feasible family; refugee preferences play essentially no role beyond stating that some locations are infeasible (i.e., eliciting dichotomous preferences is sufficient). 
Yet beyond the inefficiency of a serial dictatorship, a key reason that made MWP particularly attractive to HIAS in the context of Uniting for Ukraine, was that it gives refugees a \emph{minimal choice} of sponsor location. 
This feature was important as a way of giving refugees greater agency (objective (4)). One can imagine mechanisms in which a family could have the opportunity to consider several sponsors sequentially before being removed from the system after several rejections.\footnote{Such mechanisms are used for the allocation of public housing \cite{Thakral}.} However, such mechanisms create congestion in the system because it takes time to consult with each family and simultaneous offers would need to be mediated. In the context of rapid refugee relocation, such delays may be undesirable; moreover, such mechanisms are inefficient and create justified envy. MWP is therefore an attractive compromise: it avoids congestion (meeting objective (2)) and has good efficiency properties driven by allowing refugees to express the intensity of their preferences over locations.
Finally, HIAS was concerned that sponsors were also served well by RUTH. In particular, there were concerns that some sponsors in fewer destinations might need to wait in the system for a long time before the match (meeting Objective (2)). Furthermore, HIAS was concerned that the constraints of the sponsors were carefully taken into account and that HIAS could explain to sponsors why a particular match happened. While many sponsors signed up early on in the project, they were quickly matched and the rates of arrival varied considerably across locations. However, there are reasons to believe that several features of RUTH have had a positive effect on sponsor supply. First, sponsor constraints were faithfully taken into account by RUTH. Second, once the sponsor signed up, ``finding'' an eligible refugee family was very fast. Third, since sponsors knew that refugees' preferences were at a forefront of the process, they were incentivized to provide information about their area and potentially even improve the quality of their hosting offer.
\begin{algorithm}[t]
\caption{Multiple Waitlist Procedure as Implemented in RUTH$^*$}
\label{alg:mwp}
\begin{algorithmic}[1]
\scriptsize
  \REQUIRE main queue $Q$, locational queue $Q_{\ell}$, sponsor set $C$, family set $F$ and period  $t$ \\ $t$ initialized to $0$ \\

    \STATE $Q \gets f$ 
  \IF{$C = \emptyset$ this period $t$}
    \STATE $t=t+1$, and go to Step 1.
  \ELSE
    \STATE Randomly select sponsor $c' \in C$ with associated location $\ell'$.
  \ENDIF
  \WHILE{$Q_{\ell'} \neq \emptyset$}
    \STATE Offer $c'$ to feasible family $f' \in Q_{\ell'}$ in FIFO order.
  \IF{$f'$ accepts}
    \RETURN Match $(c',f'), C \leftarrow C \setminus c'$, and $Q_{\ell'} \leftarrow Q_{\ell'} \setminus f'$, and go to Step 1.
  \ELSE
    \STATE Remove family $f'$ from $Q_{\ell'}$ and entire system (manual match).
    \STATE Offer sponsor $c'$ to next family in $Q_{\ell'}$.
  \ENDIF
  \ENDWHILE
  \IF{$Q_{\ell'} = \emptyset$ OR $c'$ remains unmatched}
    \STATE Offer $c'$ to feasible family $\bar{f} \in Q$ in FIFO order.
    \IF{$\bar{f}$ accepts}
      \RETURN Match $(c',\bar{f}), C \leftarrow C \setminus c'$, and $Q \leftarrow Q \setminus \bar{f}$, and go to Step 1.
    \ELSE
      \STATE Move family $\bar{f}$ from $Q$ to their preferred locational queue $Q_{\bar{\ell}}$, and offer sponsor $c'$ to the next feasible family in $Q$. If there is no next feasible family in $Q$, $c'$ remains in $C$.
    \ENDIF
  \ENDIF

\end{algorithmic}
\textsuperscript{*Note: an earlier version that implemented rotating sub-queues and matched a few Ukrainian families is in Appendix A.1}
\end{algorithm}

\section{Implementation}
\begin{figure}[H]
    \centering
    \includegraphics[width=0.9\textwidth]{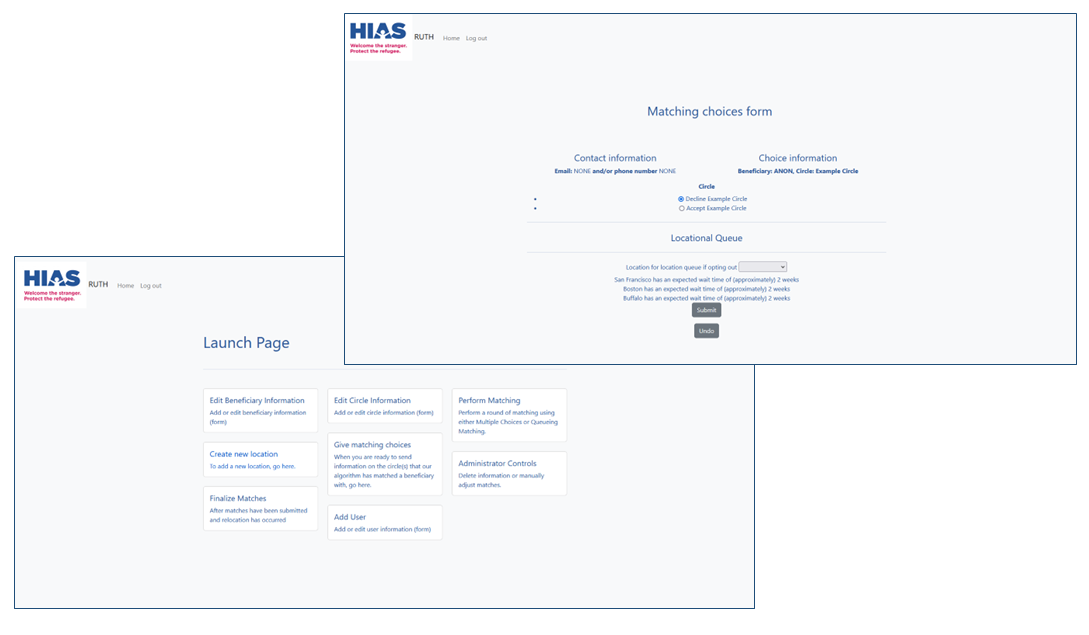}
    \caption{RUTH Interface}
    \label{pic:interface}
\end{figure}

We now describe in more detail how the MWP was implemented in RUTH, highlighting the continual evolution of the system as we learned from our experiences and made refinements along the way, culminating in the RUTH interface and pages with which HIAS staff interacts, as depicted in Figure~\ref{pic:interface}.

RUTH relies on refugee preference data that is collected through custom forms internal to the RUTH system, stored entirely on a secure HIAS server. 
The forms are completed by the primary applicant of each refugee family in Eastern Europe, with the assistance of a HIAS relocation officer.
All refugee data, including features such as working adults, religion, and medical conditions, are kept confidential on the secure HIAS server, and refugee data accessed by algorithms are deidentified to ensure the anonymity and privacy of the applicants. Refugee families were made aware of the matching algorithm and its properties and advised to reveal their preferences over locations as truthfully as possible.
Sponsor location data is also stored with RUTH and includes features such as contact information, cost of living, religion, a maximum capacity constraint on family size, the ability to support medical conditions, and a sponsor-submitted description. 


Upon the rapid initial deployment of RUTH, HIAS staff expressed the need to override some of our initial matchings and manually introduce their own selections throughout the process.
Notwithstanding compromising algorithmic guarantees, this was a critical adaptation that enabled HIAS to accommodate constraints and respond to unforeseen real-life situations requiring human intervention.
We created a dedicated control panel to address this need, empowering HIAS staff to manually delete, match, en-queue, and suspend the matching process for refugees and sponsors.

In response to observing actual matches, we adapted the queue ordering.
Our first approach involved a rotating system of four refugee family sub-queues that aimed to balance fairness and prioritize more vulnerable populations, including families with children, as well as those with medical conditions (see Appendix A).
While the initial design still satisfied the desirable properties described above\footnote{As the acyclicity assumption for Thakral's Proposition~4 was satisfied.}, it quickly became apparent that disparities in waiting times among sub-queues necessitated a transition. We thus adopted the current system that is driven by a single, main FIFO queue that prioritizes the peak waiting time over all refugees irrespective of other characteristics.

In the preference survey, it would have been unreasonable to ask for indifference-inducing waiting times between locations. Instead, we initially simply asked refugees to rank all locations that were feasible for them. However, refugee families often left survey preferences incomplete, likely due to the large number of locations.
Upon closer examination of the preference data, we determined that choices ranked lower than third-most preferred might have been poorly informed. In order to make offers in the main queue, we do not, of course, need a complete ranking anyway---we merely need to know whether or not a location is feasible.
Refugees now provide a dichotomous selection for locations they would or would not consider for placement and are only asked to rank their top three preferences. If a refugee family turns down a main queue offer, then they are only asked at that point which locational queue they want to join. This reduced the informational load on refugees, while providing us with invaluable data, as we show in Section~6. 

A key consideration underlying many of our design choices was the ability to audit when and why decisions were made.
We enable the functionality to record when refugee family matches became final, signifying their complete removal from the system, and distinct from the decision to accept the match.
Our database stores data on placement rounds, manual choices, and timestamps for refugee families and sponsor creation.
This data allows for comprehensive auditing of system outcomes, fostering the identification of potential concerns and ensuring system reliability.

The workflow that we developed for HIAS included several user features to make the implementation of RUTH from the practitioner's perspective easier to work with.
We implemented a page where HIAS staff
are able to claim matches to work on to divide and coordinate work.
They are able to see all ongoing matches, and when the matching process is executed, match information will be exported as a PDF of details to forward to the refugee family to allow them to accept or decline the proposed match.
We added a feature that tracks the status of matches in all views, allowing them to be 'frozen' (artificially removed from consideration manually), and displaying their matched dates for when their match was accepted when viewing the full list of refugees.
Finally, we added a finalized matches page to track when families actually arrive in the US.

From a software architecture standpoint, we used a SQL database to manage the data, the choice of database is independent of implementation because we used sqlalchemy as an ORM layer.
The SQL database was given an interface for end-users at HIAS by hosting it on their internal servers which were behind authentication first to connect to their internal network and then to authenticate within the context of the application.
The schema was designed to be fully normalized with tables for sponsors, locations, sub-queues (not currently in use), families, users, locational queues, matchings, friends, preferences, willingness to match, and a persisted global constant for what round of matching is used.
Logging, database connection, and forms are done by Flask (a Python web framework) with the added plugins for features in security, ORM, forms, and login.

\begin{figure}[htbp]
    \centering
    \includegraphics[width=1\textwidth]{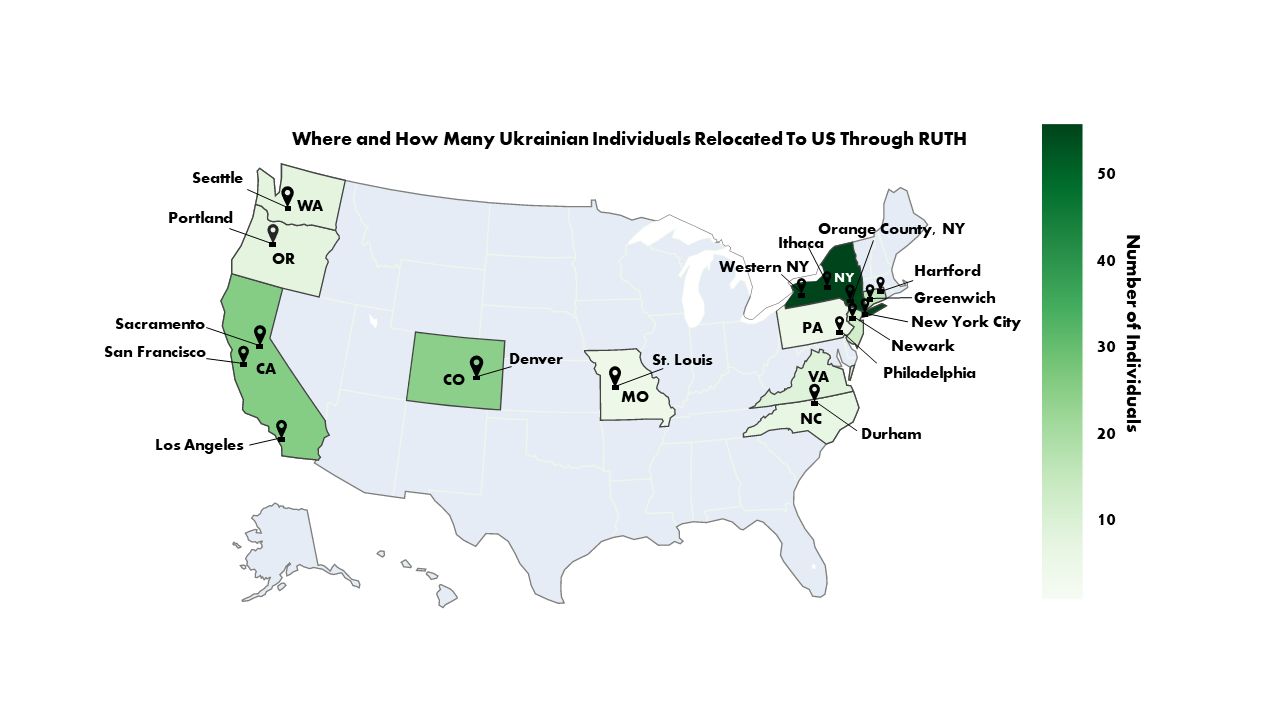}
    \caption{Heat map of individuals relocated to US. Note: Durham includes matches from nearby Halifax, VA.}
    \label{pic:matches}
\end{figure}

\section{Descriptive Analysis}
\begin{figure}[h]
    \centering
\includegraphics[width=.5\textwidth]{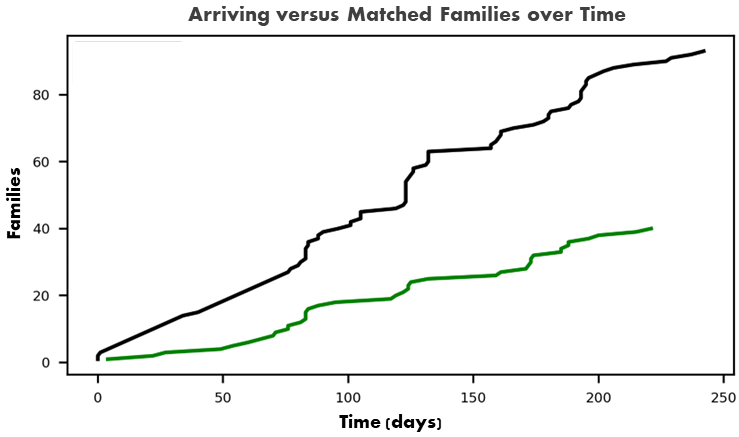}
    \caption{Number of arriving (black) and matched (green) families over time.}
    \label{pic:FamilyArrival}
\end{figure}

\begin{figure}[htbp]
    \centering
\includegraphics[width=.5\textwidth]{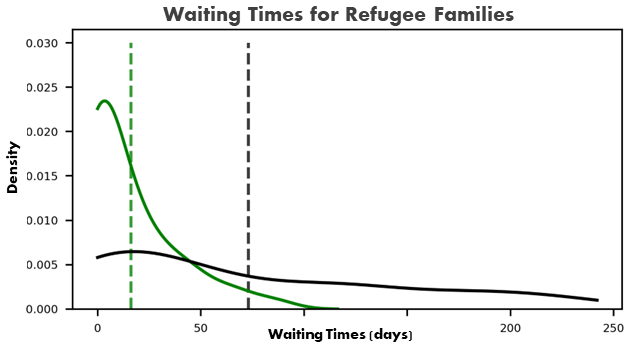}
    \caption{Distribution of waiting times for matched families (green) and all families (black)}
    \label{pic:distwaiting}
\end{figure}

We now briefly describe some basic facts from our data.
Figure~\ref{pic:matches} shows the total number of individuals relocated through RUTH, including those who were matched algorithmically by the MWP and those who were manually matched by HIAS staff. 
As of mid-April 2023, RUTH has relocated 136 individuals (40 families) with an average family size of 3.4. As Figure~\ref{pic:FamilyArrival} shows the rate of families joining the system outstripped the matching rate from the very start. The average waiting time for matched families was 16 days and for all families was 73 days. Figure~\ref{pic:distwaiting} shows that there was, however, substantial variation in waiting times for families.

While the highest concentration of matched refugees in the data is in major metropolitan cities in states such as New York, New Jersey, California, and Colorado, there are also clusters of matched refugees in smaller cities such as Greenwich, Connecticut, and Halifax, Virginia.
The data thus reveal a diverse range of matched locations that were informed by refugee preferences.
Among these matches, four families opted to join locational queues (of San Francisco and New York City).
Two of these families incurred wait times of 41 and 50 days, respectively, in order to be successfully matched with sponsors in their desired locations. 

\begin{figure}[h]
    \centering
    \includegraphics[width=0.9\textwidth]{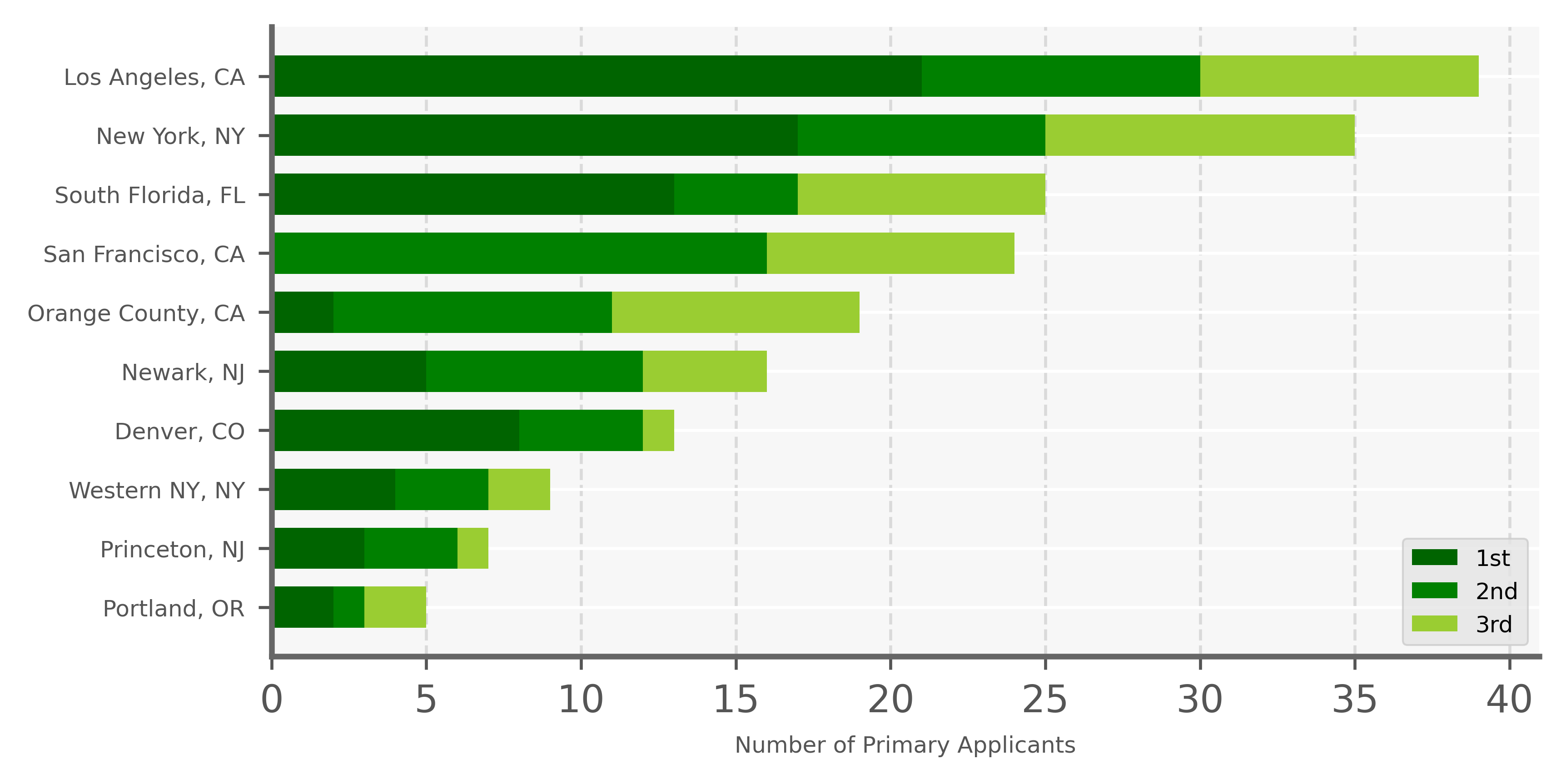}
    \caption{Observed preference distribution over sponsor locations, ordered by sums over top three ranks. Note: locations with fewer than five top rankings have been removed to ensure privacy.}
    \label{pic:PrefDist}
\end{figure}

The preference data depicted in Figure~\ref{pic:PrefDist}
is collected from over 85 families and highlights the substantial diversity in refugee preferences over locations. 
This heterogeneity reflects the private information of refugees, which may not be captured by traditional feature-based estimation methods. 
For example, some refugees may prioritize access to higher education, while others may prioritize social connections, and these preferences cannot be readily inferred from refugees observable characteristics.
Figure~\ref{pic:PrefDist} reveals that Los Angeles, New York, and South Florida are ranked among the top three choices of refugees by total rankings, with San Francisco a close fourth.
We note that Los Angeles is ranked as the top choice by the refugee population, while Sacramento, though also in California and having a known Ukrainian diaspora, was ranked much lower. 
This empirical evidence underscores the value of eliciting truthful preferences in any preference-based matching system, as it can improve the relocation process by leveraging the private information of refugees.

The demographic characteristics of refugees relocated through RUTH and those resettled through traditional refugee resettlement are presented in Table~\ref{table:demog11}. Ukrainian refugees are more middle-aged and are more likely to speak English than refugees (from all destinations) who arrived via the conventional, top-down resettlement pathway.
Like other resettled refugees, many Ukrainians had already fled Ukraine to nearby countries, such as Poland (32\%) and Germany (16\%). However, some Ukrainians had not crossed an international border and applied, often as internally displaced persons, directly from Ukraine (21\%).

\begin{center}
\begin{tabular}{p{4cm}|p{3cm}|p{3cm}}
\toprule
\textbf{Applicant Features} & \textbf{RUTH} & \textbf{Resettlement} \\
\midrule
Male & 49\%& 53\%\\
Family Size & 3.4 & 2.55\\
English-speaking & 76\% & 42\% \\
Age: & & \\
\hspace{0.5cm}18-29 & 18\% & 45\% \\
\hspace{0.5cm}30-39 & 46\% & 28\% \\
\hspace{0.5cm}40-49 & 25\% & 16\% \\
\hspace{0.5cm}$>$50 & 11\% & 11\% \\

\multicolumn{3}{p{10cm}}{\footnotesize{Resettlement considers data on refugee resettlement to the United States for the period from quarter 1, 2011 to quarter 3, 2016~\cite{ban18}. Family size data for 2017 is from \cite{ahani21}. Features relate to primary applicants.}}

\end{tabular}
\captionof{table}{Distribution of applicant features.}
\label{table:demog11}
\end{center}

\vspace{-5mm}

\section{Counterfactual Analysis}
We now report results from two counterfactual analyses: the implementation of periodic Top Trading Cycles and tuning long-run steady state arrival rates of sponsors across locations in order to maximize matching rates.
\subsection{Periodic Top Trading Cycles}
In the RUTH implementation of the MWP, we try to match a sponsor to a refugee as soon as possible. This results in an inefficiency: a refugee might miss out on a sponsor merely because they happened to have joined the system too late or too early. One option to eliminate this type of inefficiency is to wait for all the sponsors and all the refugees to arrive and match them all at once. One candidate for matching many refugees and sponsors at once is the Top Trading Cycles (TTC) algorithm~\cite{SHAPLEY1974}, which we discussed with HIAS early in the project as a possible candidate that could meet their objectives (see Section~\ref{sec:vsobj}). TTC is a strategy-proof mechanism and, if waiting times are ignored, it is also ex-post efficient.\footnote{We considered two further options. First, because this is a one-to-one matching problem, we could have also used the refugee-proposing deferred acceptance algorithm, as we know that no individually rational matching is strictly preferred by all refugees to the refugee-optimal stable matching. Second, we consider a serial dictatorship. However, deferred acceptance is inefficient in the case where sponsors might have been able to host more than one family and serial dictatorship could not cope with potentially heterogeneous priorities or locations/sponsors. TTC is efficient when sponsors can host multiple families and can deal with heterogeneous priorities, although so far, neither of these features has been incorporated into RUTH.} However, using TTC infrequently can be highly impractical because the waiting times for sponsors and refugees might become unacceptably high. Nevertheless, one could imagine running TTC periodically, e.g., every week or every month, with all the families and sponsors that arrived during that period and that were left over from previous ones.

Let us briefly review how the TTC algorithm works. We represent families and sponsors as nodes of a graph.
Each unmatched family ``points'' with a directed edge at its most preferred sponsor (we explain below how we construct preferences over sponsors) and each sponsor points at the highest-priority feasible family. There must be at least one cycle; in each cycle, match the family to the sponsor it is pointing at. Remove matched families and sponsors and ask the remaining families and sponsors to point again until no more pointing occurs.
 
In RUTH, we match families with sponsors while families express their preferences over sponsor locations, causing ties among sponsors affiliated with a common location. In consultation with HIAS staff, we adopted varying tie-breaking rules informed by available data and expert knowledge to assure heterogeneous preferences.   
Our approach involves ranking sponsors within the same location-based first on religion and then on housing affordability. 
Similarly, sponsor priorities over families consider religious affiliation, and in the case of homogeneous families, we distinguish them by family size to comply with TTC requiring heterogeneous preferences to find a cycle for each stable match, with any further ties broken arbitrarily~\cite{SHAPLEY1974}.        

We examined counterfactual scenarios involving the use of TTC at various intervals for families and sponsors. Our approach employs a batch-style matching process, which can afford HIAS control over the matching frequency at the expense of waiting time accrued by refugee families.
To do this, we define a notion of a normalized average rank that captures match quality.
The normalized average rank is defined as the average over matches of the probability that if a location is chosen at random from among those the family ranked, the new location will have a better ranking.
For example, if a family has ranked 5 locations and was matched to its second choice, the normalized rank is 0.2, indicating that 1/5 of the choices would have performed better.
The normalized average rank, therefore, adjusts for the fact that some families ranked different numbers of locations than other families.

\begin{figure}[t]
    \centering
    \includegraphics[width=0.5\textwidth]{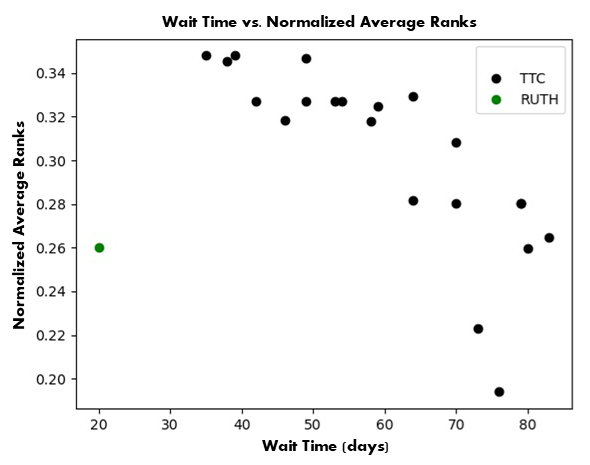}
    \caption{The trade-off between wait time and match quality, plotted for RUTH (green) and 20 runs of TTC (black).}
    \label{pic:TTC}
\end{figure}

Figure~\ref{pic:TTC} shows that there is a trade-off between the waiting time and normalized rank of locations in our data.
Increasing the frequency of TTC runs decreases the overall match quality yet reduces the waiting time for families. We find that with a periodic Top Trading Cycles algorithm, increasing the period length from 24 days to 80 days (corresponding to increasing the waiting times of matched families from 47 days to 83 days) improves the normalized average rank of the match from 0.38 to 0.26. 
Figure~\ref{pic:TTC} also shows that RUTH substantially outperforms periodic TTC given the match quality it achieves. To achieve the normalized rank produced by RUTH, periodic TTC induces waiting times that are \emph{four} times longer.  



\subsection{Optimizing Sponsor Arrival Rates}
We now consider which levels of sponsor arrival rates in each location are consistent with a long-run process of refugee arrival. Understanding what the arrival rates of sponsors need to be in order to quickly clear queues can help the refugee agency with allocating resources for sponsor recruitment. In order to estimate the necessary arrival rates, we simulate an arrival process of refugees based on our data and adjust sponsor arrival rates to ensure that queues at each location remain roughly constant over time.

\subsubsection{Simulation setup}
The model we choose to simulate starts with two arrival processes, both empirically informed by the arrival rate and characteristics of families into RUTH. We use a Poisson arrival process for both families and sponsors with a rate of $93$ families (or sponsors) in $242$ days. A key step in the simulation is to model differences in preferences over the locations. 
To capture the decision process of refugees in the MWP, we employ a Random Utility Model (RUM) to model the latent preferences of families and their choices regarding accepting or rejecting a main queue match, as well as the selection of a locational queue to join (see Appendix D.2 for more details).


\subsubsection{Results} We consider two cases of sponsor arrival rates. In the \emph{uniform} case, each location has equal sponsor arrival rates. In the \emph{tuned} case, the sponsor arrival rates reflect how much refugee families ``demand'' each location---this is revealed from their long-run choices of locational queues.\footnote{To understand the drift in the steady-state behavior of the uniform case compared with the tuned case, see Appendix D.2.1.}



\begin{figure}[h]
    \centering
    \includegraphics[width=0.6\textwidth]{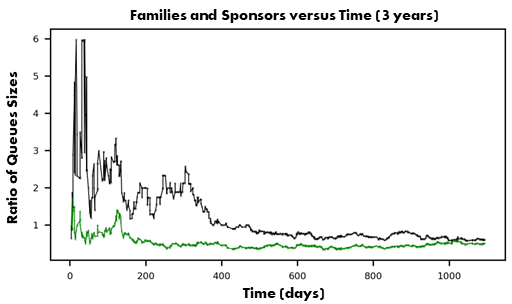}
    \caption{Tuned scenario queue size over uniform scenario queue size for unmatched families (black) and sponsors (green) over 3 years.}
    \label{pic:Ratio-size}
\end{figure}

\begin{figure}[h]
    \centering
    \includegraphics[width=0.8\textwidth]{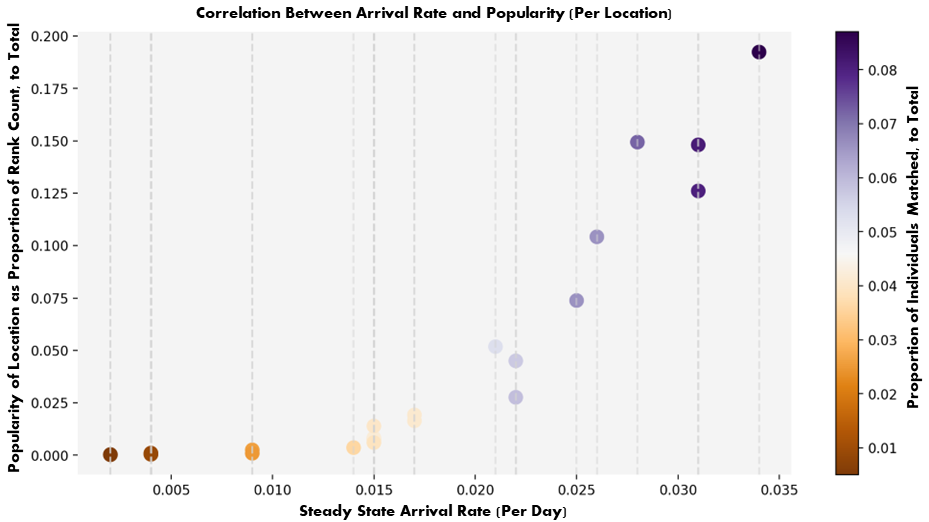}
    \caption{Estimated arrival rate of sponsors in each location in the steady state of the tuned scenario versus preference popularity as in Figure~\ref{pic:PrefDist}.}
    \label{pic:ArrivalSimulation}
\end{figure}

Figure~\ref{pic:Ratio-size} shows how queue sizes (for refugees and for sponsors) evolve over time in the uniform versus tuned scenario. A ratio less than one suggests that the queue under the tuned scenario is shorter.
As Figure~\ref{pic:Ratio-size} shows, queue lengths in the tuned scenario relative to the uniform scenario fall within a few months of simulation. As a result, the tuned scenario matches far more refugees in the long run. Figure~\ref{pic:ArrivalSimulation} shows that the locations where sponsor arrival rates are highest in the tuned scenario are the ones that are highly ranked by the refugee families in the simulation\footnote{In Appendix D.2.1, we show that these patterns also hold when using the empirical preference distribution, see Figure 12}.
This suggests that there may be substantial value for the refugee agency to base investment decisions for sponsor recruitment on refugee preferences.

\section{Conclusion} 
This paper introduced RUTH, an algorithmic matching system for Ukrainians seeking sponsored humanitarian parole in the United States. RUTH is implemented and deployed via a custom-built, secure web app in partnership with HIAS, a major refugee resettlement agency. To match refugees to sponsors, RUTH uses \citeauthor{Thakral}'s~(\citeyear{Thakral}) Multiple Waitlist Procedure augmented with feasibility factors such as family size, medical needs, and religious practice compatibility.

We showed that refugee preferences over locations are diverse and as a result, their proper incorporation into the parole process might have benefits for refugees and hosts. The performance of RUTH in our setting compares favorably to periodic Top Trading Cycles and insights gleaned from RUTH can help refugee agencies plan for investment in recruitment of sponsors across different locations.

Our work opens up many directions for further research and practice.
On the research side, one could: (i) benchmark RUTH against other queuing algorithms in the literature; (ii) investigate to what extent features of locations explain the preferences of refugees; (iii) study the impact of incorporating refugee preferences on the measurable outcomes (such as employment) of refugees. 
On the practice end, one could: (i) adapt RUTH to work in other humanitarian parole processes; (ii) investigate the effect of different inputs into the MWP (such as refugee priority) on outcomes; (iii) experiment with creative and cost-effective ways of collecting refugee preferences and other data.

The early success of RUTH in the Ukrainian refugee crisis suggests the possibility for broader adoption of preference-based algorithms in other refugee situations worldwide.
By incorporating refugee preferences and sponsor priorities, RUTH promises a more humane and fairer future for both refugees and host communities.

In the long term, this research could fundamentally change the paradigm of refugee resettlement. By acknowledging and integrating the preferences of refugees into the matching process, RUTH offers a fresh, human-centered approach that contrasts with current one-size-fits-all solutions. Beyond the immediate potential to improve refugee satisfaction and integration outcomes, such a shift could have profound societal implications. First, individuals matched to locations that align with their needs and preferences could potentially form stronger community ties, secure more meaningful employment, and make additional positive contributions to their new communities. Second, RUTH's success can provide a versatile model for other crucial matching scenarios, such as job placements and housing programs, underscoring the wide-ranging utility of such preference-based algorithms. Moreover, the data-driven insights yielded by RUTH can inform more equitable and strategic policy-making. Specifically, the nuanced understanding of refugee preferences and determining factors can direct resource allocation in areas like language learning, job training, and community development, thereby aligning investments with the true needs and aspirations of refugees.

\section*{Acknowledgement}
We have been privileged to collaborate with HIAS and want to especially thank Alicia Wrenn, Chloe Shiras, Isabel Burton and Ellen Kotik for their continued partnership.
We thank Neil Thakral for helpful conversations that have improved the manuscript.
We are also grateful for the support of the National Science Foundation (Operations Engineering) RAPID award CMMI-2233377.

\bibliographystyle{unsrtnat}
\bibliography{Lit}

\begin{thebibliography}{28}
\providecommand{\natexlab}[1]{#1}
\providecommand{\url}[1]{\texttt{#1}}
\expandafter\ifx\csname urlstyle\endcsname\relax
  \providecommand{\doi}[1]{doi: #1}\else
  \providecommand{\doi}{doi: \begingroup \urlstyle{rm}\Url}\fi

\bibitem[Thakral(2016)]{Thakral}
Neil Thakral.
\newblock The public-housing allocation problem: Theory and evidence from {Pittsburgh}.
\newblock 2016.

\bibitem[UNHCR(2023{\natexlab{a}})]{UNHCR1}
UNHCR.
\newblock {Six faces of the forcibly displaced in Ukraine}.
\newblock Accessed May 9, 2023, \url{https://www.unhcr.org/ua/en}, 2023{\natexlab{a}}.

\bibitem[UNHCR(2023{\natexlab{b}})]{UNHCR2}
UNHCR.
\newblock {Ukraine Emergency}.
\newblock Accessed May 9, 2023, \url{https://www.unhcr.org/emergencies/ukraine-emergency}, 2023{\natexlab{b}}.

\bibitem[NBCNews(2023)]{Over100}
NBCNews.
\newblock { U.S. has admitted 271,000 Ukrainian refugees since Russian invasion, far above Biden’s goal of 100,000}.
\newblock \url{https://www.nbcnews.com/politics/immigration/us-admits-271000-ukrainian-refugees-russia-invasion-biden-rcna72177}, 2023.

\bibitem[{\r A}slund and Rooth(2007)]{aaslund2007and}
Olof {\r A}slund and Dan-Olof Rooth.
\newblock Do when and where matter? {Initial} labour market conditions and immigrant earnings.
\newblock \emph{The Economic Journal}, 117\penalty0 (518):\penalty0 422--448, 2007.

\bibitem[{\r A}slund et~al.(2010){\r A}slund, {\"O}sth, and Zenou]{aaslund2010important}
Olof {\r A}slund, John {\"O}sth, and Yves Zenou.
\newblock How important is access to jobs? {Old} question, improved answer.
\newblock \emph{Journal of Economic Geography}, 10\penalty0 (3):\penalty0 389--422, 2010.

\bibitem[Damm(2014)]{damm2014neighborhood}
Anna~Piil Damm.
\newblock Neighborhood quality and labor market outcomes: Evidence from quasi-random neighborhood assignment of immigrants.
\newblock \emph{Journal of Urban Economics}, 79:\penalty0 139--166, 2014.

\bibitem[Bansak et~al.(2018)Bansak, Ferwerda, Hainmueller, Dillon, Hangartner, Lawrence, and Weinstein]{ban18}
Kirk Bansak, Jeremy Ferwerda, Jens Hainmueller, Andrea Dillon, Dominik Hangartner, Duncan Lawrence, and Jeremy Weinstein.
\newblock Improving refugee integration through data-driven algorithmic assignment.
\newblock \emph{Science}, 359\penalty0 (6373):\penalty0 325--329, 2018.

\bibitem[Bansak and Paulson(2022)]{bansak2022dynamic}
Kirk Bansak and Elisabeth Paulson.
\newblock Outcome-driven dynamic refugee assignment with allocation balancing.
\newblock In \emph{Proceedings of the 23rd ACM Conference on Economics and Computation}, pages 1182--1183, 2022.

\bibitem[Ahani et~al.(2021)Ahani, Andersson, Martinello, Teytelboym, and Trapp]{ahani21}
Narges Ahani, Tommy Andersson, Alessandro Martinello, Alexander Teytelboym, and Andrew~C. Trapp.
\newblock Placement optimization in refugee resettlement.
\newblock \emph{Operations Research}, 69\penalty0 (5):\penalty0 1468–--1486, 2021.

\bibitem[Ahani et~al.(2023)Ahani, G{\"o}lz, Procaccia, Teytelboym, and Trapp]{ahani23}
Narges Ahani, Paul G{\"o}lz, Ariel~D Procaccia, Alexander Teytelboym, and Andrew~C Trapp.
\newblock Dynamic placement in refugee resettlement.
\newblock \emph{accepted, Operations Research}, 2023.

\bibitem[Jones and Teytelboym(2017)]{jon16}
Will Jones and Alexander Teytelboym.
\newblock The local refugee match: Aligning refugees' preferences with the capacities and priorities of localities.
\newblock \emph{Journal of Refugee Studies}, 31:\penalty0 152--178, 2017.

\bibitem[Delacr{\'e}taz et~al.(2020)Delacr{\'e}taz, Kominers, and Teytelboym]{delacretaz2019matching}
David Delacr{\'e}taz, Scott~Duke Kominers, and Alexander Teytelboym.
\newblock {Matching mechanisms for refugee resettlement}.
\newblock \emph{American Economic Review}, 2020.

\bibitem[{U.S. Department of State}(2023{\natexlab{a}})]{wraps2023}
{U.S. Department of State}.
\newblock Refugee arrivals by state and nationality as of may 9, 2023.
\newblock Accessed May 9, 2023, \url{https://www.wrapsnet.org/admissions-and-arrivals}, 2023{\natexlab{a}}.

\bibitem[{U.S. Department of State}(2023{\natexlab{b}})]{StateRefugeeAdmissions}
{U.S. Department of State}.
\newblock Refugee admissions.
\newblock Accessed May 9, 2023, \url{https://www.state.gov/refugee-admissions}, 2023{\natexlab{b}}.

\bibitem[{Migration Policy Institute}(2022)]{MPITrump}
{Migration Policy Institute}.
\newblock Four years of profound change: Immigration policy during the trump presidency.
\newblock Accessed May 9, 2023, \url{https://www.migrationpolicy.org/research/four-years-change-immigration-trump}, 2022.

\bibitem[{U.S. Citizenship and Immigration Services}(2023)]{USCISHP}
{U.S. Citizenship and Immigration Services}.
\newblock Humanitarian parole.
\newblock Accessed May 9, 2023, \url{https://www.uscis.gov/forms/explore-my-options/humanitarian-parole}, 2023.

\bibitem[Freund et~al.(2023)Freund, Lykouris, Paulson, Sturt, and Weng]{freund2023group}
Daniel Freund, Thodoris Lykouris, Elisabeth Paulson, Bradley Sturt, and Wentao Weng.
\newblock Group fairness in dynamic refugee assignment, 2023.

\bibitem[Jones and Teytelboym(2018)]{jones2018local}
Will Jones and Alexander Teytelboym.
\newblock The local refugee match: Aligning refugees’ preferences with the capacities and priorities of localities.
\newblock \emph{Journal of Refugee Studies}, 31\penalty0 (2):\penalty0 152--178, 2018.

\bibitem[Andersson and Ehlers(2020)]{andersson2020assigning}
Tommy Andersson and Lars Ehlers.
\newblock Assigning refugees to landlords in sweden: Efficient, stable, and maximum matchings.
\newblock \emph{The Scandinavian Journal of Economics}, 122\penalty0 (3):\penalty0 937--965, 2020.

\bibitem[Acharya et~al.(2022)Acharya, Bansak, and Hainmueller]{acharya2022combining}
Avidit Acharya, Kirk Bansak, and Jens Hainmueller.
\newblock Combining outcome-based and preference-based matching: A constrained priority mechanism.
\newblock \emph{Political Analysis}, 30\penalty0 (1):\penalty0 89--112, 2022.

\bibitem[Caspari(2018)]{Caspari}
G.~Caspari.
\newblock An alternative approach to asylum assignment.
\newblock \emph{Bonn University Working Paper}, 2018.

\bibitem[Andersson et~al.(2018)Andersson, Ehlers, and Martinello]{TommyAnderssonD}
Tommy Andersson, Lars Ehlers, and Alessandro Martinello.
\newblock Dynamic refugee matching.
\newblock \emph{Working Paper}, 2018.

\bibitem[Leshno(2022)]{Jacob22}
Jacob~D. Leshno.
\newblock Dynamic matching in overloaded waiting lists.
\newblock \emph{American Economic Review}, 112\penalty0 (12):\penalty0 3876--3910, 2022.

\bibitem[Doval(2022)]{doval2022dynamically}
Laura Doval.
\newblock {Dynamically stable matching}.
\newblock \emph{Theoretical Economics}, 17\penalty0 (2):\penalty0 687--724, 2022.

\bibitem[Bloch and Cantala(2017)]{BlochCantala}
Francis Bloch and David Cantala.
\newblock Dynamic assignment of objects to queuing agents.
\newblock \emph{American Economic Journal: Microeconomics}, 9\penalty0 (1):\penalty0 88--122, 2017.

\bibitem[Arnosti and Shi(2020)]{ArnostiShi}
Nick Arnosti and Peng Shi.
\newblock Design of lotteries and wait-lists for affordable housing allocation.
\newblock \emph{Management Science}, 66\penalty0 (6):\penalty0 2291--2799, 2020.

\bibitem[Shapley and Scarf(1974)]{SHAPLEY1974}
Lloyd Shapley and Herbert Scarf.
\newblock On cores and indivisibility.
\newblock \emph{Journal of Mathematical Economics}, 1\penalty0 (1):\penalty0 23--37, 1974.
\newblock \doi{https://doi.org/10.1016/0304-4068(74)90033-0}.

\end{thebibliography}
\newpage
\appendix
\section*{Appendix}

\section{Evolution of RUTH} \label{sec:mwp_1}
\subsection{Using subqueues in the priority order of the MWP} \label{App_UsingSubQueues}
The Multiple-Waitlist Procedure (MWP), introduced by~\citet{Thakral}, was initially implemented with rotating $Sub_{Q}$ where families were categorized based on having medical needs, friends in the US, children, and others. The pseudocode is given in Algorithm~\ref{alg:mwp_1}.

\begin{algorithm}[H]
\caption{Multiple Waitlist Procedure as Implemented in RUTH (Initial Version)}
\label{alg:mwp_1}
\begin{algorithmic}[1]
\scriptsize
  \REQUIRE main queue $Q$, subqueue $Sub_{Q}$, locational queue $Q_{\ell}$, sponsor set $C$, family set $F$ and period $t$  \\ $t$ initialized to $0$ \\

    \STATE $Sub_{Q} \gets f$ 
  \IF{$C = \emptyset$ this period $t$}
    \STATE $t=t+1$, and go to Step 1.
  \ELSE
    \STATE Randomly select sponsor $c' \in C$ with associated location $\ell'$.
  \ENDIF
  \WHILE{$Q_{\ell'} \neq \emptyset$}
    \STATE Offer $c'$ to feasible family $f' \in Q_{\ell'}$ in FIFO order.
  \IF{$f'$ accepts}
    \RETURN Match $(c',f'), C \leftarrow C \setminus c'$, and $Q_{\ell'} \leftarrow Q_{\ell'} \setminus f'$, and go to Step 1.
  \ELSE
    \STATE Remove family $f'$ from $Q_{\ell'}$ and entire system (manual match).
    \STATE Offer sponsor $c'$ to next family in $Q_{\ell'}$.
  \ENDIF
  \ENDWHILE
  \IF{$Q_{\ell'} = \emptyset$ OR $c'$ remains unmatched}
    \STATE Offer $c'$ to feasible $\bar{f} \in Sub_{Q}$.
    \IF{$\bar{f}$ accepts}
      \RETURN Match $(c',\bar{f}), C \leftarrow C \setminus c'$, and $Q \leftarrow Q \setminus \bar{f}$, and go to Step 1.
    \ELSE
      \STATE Move family $\bar{f}$ from $Sub_{Q}$ to the preferred locational queue $Q_{\bar{\ell}}$, and offer sponsor $c'$ to the top feasible family in next $Sub_{Q}$.
    \ENDIF
  \ENDIF

\end{algorithmic}
\end{algorithm}
\subsection{Collection of refugee preferences}
We also made changes to preference collection over time.
While collecting preference data for the first time was a valuable undertaking, this information was primarily used to check feasibility of each match based on whether the sponsor-associated location was on the refugee preference list. Later, it was determined that only the top choices were not arbitrary so with the advice of HIAS staff and earlier data observations, only the top three choices were collected. Along with this change, refugees were also asked to provide their preferred and nonpreferred locations for being matched with a host, which were regarded as feasibility criteria before presenting a match offer to the refugee family.

\section{Additional Information for Descriptive Analyses} \label{App_Extra_Descriptive}

Our intended analysis required preprocessing of the empirical data. This section details how we cleaned the data for analysis.
\subsection{Missing data on arrival} \label{App_MissingDataOnArrival}
The original design of RUTH lacked a feature for logging entry times, so the system utilized the order of family IDs to infer the sequence of refugee arrivals for queuing.
We estimated missing creation dates using linear interpolation, drawing from the start date of the first match through to the date the data was extracted for analysis, mid-April 2023, applicable to both sponsors and families.
\subsection{New location and locations without sponsors} \label{App_NewLocations}
The gradual and uncertain arrival of sponsors has led to the emergence of new locations that were not initially included in preference lists for refugees. These locations were added to the end of the preference list to provide unmatched refugees with the opportunity to obtain a first match while waiting in the main queue. In other cities, it was anticipated that sponsors would arrive in certain locations such as Western New York, Northern New York, New York City, Los Angeles, San Francisco, South Florida, and other major US cities, thus they were included from the onset of RUTH for preference collection purposes. In some locations in Figure~\ref{pic:PrefDist},there had been no sponsors throughout our data collection period, so they are not represented in Figure~\ref{pic:matches} as there were no matched refugees in those locations.
\section{Difficulty in predicting preferences using family characteristics} \label{App_DifficultyInPredictingPreferences}
We briefly describe the connection between refugee characteristics and their preferences.
Our findings suggest that these characteristics do not necessarily predict the inclusion of a family in the top three ranks.
As traditional ANOVA methods prove inadequate, and the task of predicting full rank orders is overly complex given the limited family sample size, we instead focus on a more manageable problem: employing logistic regression to predict a family's presence in the top three locations, using separate regression models for each location.

\begin{figure}[h]
    \centering
    \includegraphics[width=0.6\textwidth]{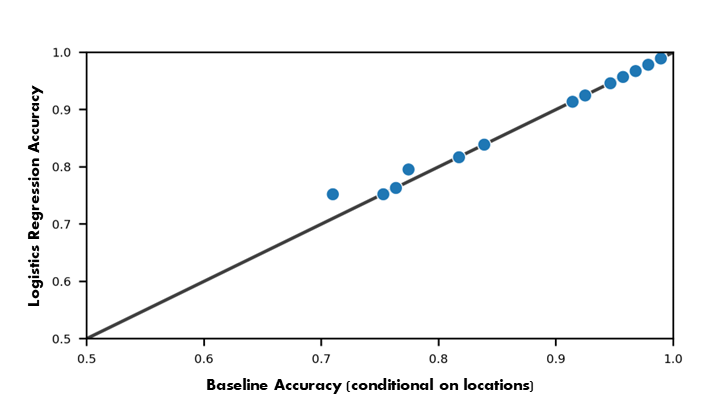}
    \caption{Accuracy comparison conditioning on all family characteristics vs. unconditional.}
    \label{pic:TopThreeAcc}
\end{figure}

If family characteristics are highly predictive of their top three preferred locations for resettlement, our predictive models ought to exhibit very high accuracy for every location.
As can be seen in Figure~\ref{pic:TopThreeAcc}, there is no improvement in the accuracy of our models over the baseline, which attempts to predict the most likely outcome without considering family characteristics.
This remains so even with our limited dataset and without a separate validation set.
This suggests that strictly classifying the top preferences of refugees based only on family characteristics may not be readily achievable and that there is no strong link exists in our data between family characteristics and location preferences.
Individual differences may play a significant role in shaping these preferences, indicating a complexity beyond the general characteristics we analyzed.

\section{Additional Information for Counterfactual Analyses}
\label{App_Extra_Counterfactual}

The remainder of the Appendix provides additional implementation details for TTC and simulation counterfactuals.

\subsection{Additional Details for Periodic Top Trading Cycles}
\label{App_Extra_TTC}

The process of matching families to locations that appeared after refugees had arrived presented a challenge due to the absence of preference data for these locations.
We limited the execution of TTC to matches with explicit preference lists, including refugees who had ranked their top three choices and earlier joined families who had expressed their preferences over all existing locations at the time.
As expected, a reduced number of ranked locations for families who only ranked their top three choices decreased the likelihood of being selected by sponsors, resulting in lower probabilities of matches occurring with TTC compared to MWP.
We also removed all matches that were completed manually, resulting in only 13 families who had a complete ranking over locations available for a comparative analysis of performance.
Thus, we compared only the matches generated in RUTH with actual existing preferences to the matches proposed by periodic TTC, which performs matching on a variable number of families depending on the batching process.

\subsection{Additional Details for Optimizing Sponsor Arrival Rates}
\label{App_Extra_Sim}

We only simulated arrivals in locations where at least one family had ranked the location within at least their top three choices to facilitate a valid comparison with the empirical distribution. 
These locations correspond to those depicted in Figure~\ref{pic:PrefDist} and other locations that are excluded from Figure~\ref{pic:PrefDist} due to the existence of fewer family rankings to maintain privacy considerations.
In our discrete event simulation process, the arrival rates of sponsors and families are equivalent, as the matches that occur in MWP during each period eliminate exactly one family and one sponsor.
Thus, the waiting queue grows linearly in expectation with time when one rate exceeds the other.
We simulate independent draws from the empirical unconditional distributions of religions, family sizes, and medical needs for families.
Similarly, for sponsors, we draw from the empirical unconditional distributions of religions, capacity, and medical support.
The arrival rate distribution for different locations is a hyperparameter of the simulation tuned in Figure~\ref{pic:Ratio-size}, mirroring our assumption of sponsor homogeneity.
The family and sponsor arrival rates are 93 per 242 days, where 93 is the number of families in RUTH in the 242 days over which we collected our data.

\begin{figure}[h]
    \centering
    \includegraphics[width=0.7\textwidth]{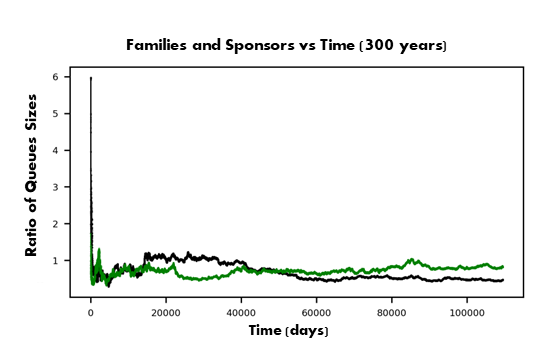}
    \caption{A ratio for tuned to uniform arrival rate of unmatched families and sponsors for 300 years, number of sponsors (green) and number of families (black).}
    \label{pic:Ratio-size2}
\end{figure}

Our model incorporates four types of constraints in the matching process, including religious affiliation compatibility, medical needs, and sufficient capacity to support family size.  
For constraints imposed by families on sponsors and religious constraints imposed by sponsors on families, the probability that a feasible sponsor arrives must exceed the probability that the constraint is imposed; otherwise, the number of families in the simulation will grow at least as fast as the arrival rate times by the difference in probabilities.
In our simulation, we chose to model constraints as independent of all other refugee family characteristics except for the aspect in which they are constraining.
While size and medical constraints are independent, religious constraints are dependent on religion, as we draw the characteristic and the constraint imposed on whether refugees (sponsors) are willing to match with the same religious affiliation.
Therefore, if the empirical data does not have this necessary property of being sufficiently unconstrained, we must adjust the probabilities, and thereby our priorities in recruiting new sponsors.
We performed necessary adjustments to satisfy these conditions, which mainly was capacity adjustment to find a steady state of arrival rates (shown in Figure~\ref{pic:ArrivalSimulation}). The impact of tuned to uniform arrival rates on number of families (sponsors) waiting in the system and queue sizes ratio are shown in~Figure~\ref{pic:Ratio-size2} and Figure\ref{pic:fam-sponsorQueue}.
The empirical data shows the same results as simulated system, depicted in Figure~\ref{pic:ArrivalSimulation}, since the majority of families and sponsors do not impose religious constraints Figure~\ref{pic:SimulationValidation}.  
By inspecting these probabilities, it is apparent that this holds for all constraints outside of sponsor capacity.
To rectify this situation, the rates of capacities have to be adjusted to meet this property, which also informs a policy of explicitly looking for sponsors with larger capacities to meet the underserved demand from large families.

We developed a model to capture families preferences over locations as follows.
Our Random Utility Model (RUM) represents the mental process of families when expressing their preference-based decisions in two steps.
First, families indicate which locations they are willing to match with at all, representing as hard constraints rather than being considered as a ranking list dependent on characteristics of families themselves. We model this process using a set of independent Bernoulli draws conditioned on families independent characteristics, represented by a per-location logistic regression. Then, from those locations, families determine an unobserved utility that they assign to each. We attempted to use the observed partial rankings directly, but realized it was unfeasible due to insufficient data and an unclear method of combining differently constrained rankings to estimate such a preference distribution. Therefore, we substituted willingness-to-match (hard constraint) logits for the utility of a location conditioned on family characteristics. This unobserved utility is used when a match arrives for the family, and they decide whether to accept immediately or join a locational queue. As such, the total utility for selecting a given locational queue is a comparison of the logits for where they have matched with a weighted combination of logits, waiting time, and estimation noise. Our utility equation is shown in Equation~\ref{eq:wtm}.

The probability with which we make a simulated family be willing to match with a location is modeled by the logistic relationship informed by $\beta$ fit using logistic regression on the actual data:
\begin{equation}
    P(\text{willing}_i|x;\beta) = \sigma (\beta^i_1 + {\beta^i_{>1}}^\top x).
    \label{eq:wtm}
\end{equation}

From there, we use the same logistic regression model to inform our estimate of the popularity of matching for the RUM process.
The utility assigned to a location for which a family does not have a match is:
\begin{equation}
    U(i|x,\beta,\rho,\lambda) = \rho_Q\times\frac{|Q_{\ell^i|}}{\lambda_i} + \rho_P\times\log(P(\text{willing}_i|x;\beta)) + g_i.
    \label{eq:wtm}
\end{equation}
While the utility assigned for a location with zero wait time is simply:
\begin{equation}
    U(i|x,\beta,\rho,\lambda) = 0 + \rho_P\times\log(P(\text{willing}_i|x;\beta)) + g_i.
    \label{eq:wtm}
\end{equation}
In this model, $\rho$ are parameters fit to maximize the parametric likelihood of 4 waitlist entries given the characteristics of the families which joined the wait list, and 36 immediate acceptances;
$\lambda$ is the arrival rate hyperparameter, dictating the expected waiting time in each queue given the number of families currently in the queue;
$g$ are random draws from the standard Gumbel distribution which is the natural noise model for a categorical decision;
$\beta$ is the result of the previous logistic regression and is used to determine logits, the intuition being that an exponential relationship exists between popularity and willingness to wait.
Refugees will maximize their sampled utility function to determine which decision they end up making in the simulation.

We estimated coefficients weighing logits, waiting time, and noise ($2-$d.o.f) by fitting a linear model to the empirical distribution of 36 accepted matches and 4 rejected matches, attempting to match the probability of rejecting a match given empirically estimated waiting times and the output of our logistic regression models. Due to the matching process, the expected absolute difference between these quantities scales as $O(\sqrt{T})$, making it impossible to achieve constant queue lengths without adjusting the arrival rates to the system's state. Nevertheless, this approach provides a useful means of comparing priorities over locations. If a constraint is satisfied less frequently than it is generated, the difference is forced to scale with $O(T)$. Thus, we call $O(\sqrt{T})$-scaling "steady-state" because it is a natural Brownian drift that is expected to occur.
Observe that not all steady-states are created equal; some will have slower random drift because the location constraints are more frequently satisfied.

\subsubsection{Additional simulation results} \label{sec:extraSimResults}

\begin{figure}[h]
    \centering
    \includegraphics[width=0.98\textwidth]{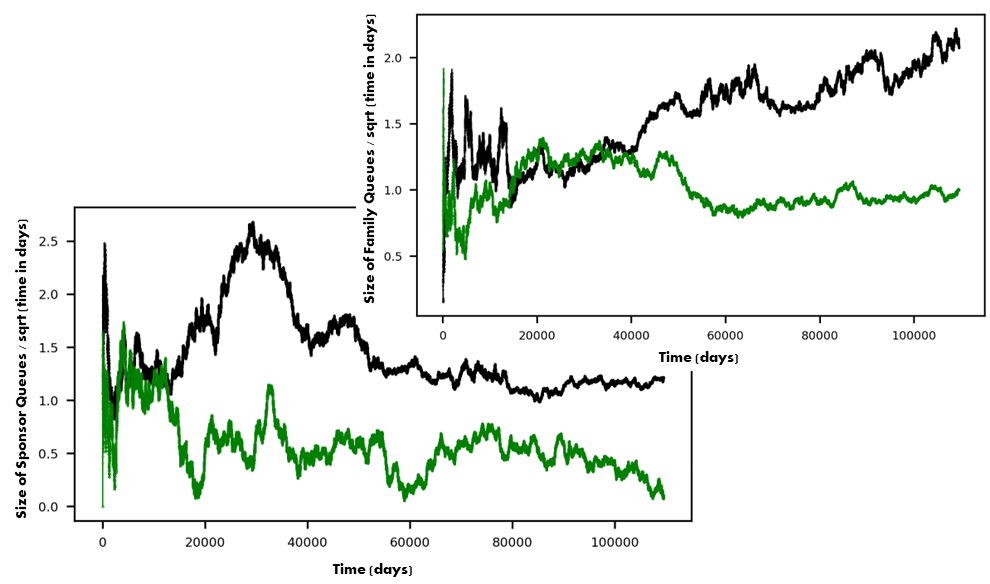}
    \caption{Long-run impact of arrival rate on size of queues, tuned (green) and uniform (black).}
    \label{pic:fam-sponsorQueue}
\end{figure}

From Figure~\ref{pic:fam-sponsorQueue}, we can see that in the long run, the system reaches a steady state for the optimized case, but when we are less careful to match the distribution of preferences, the system may be in a (slow) non-steady configuration which could grow linearly in the number of waiting sponsors over time.
This shows that there is a clear long-run benefit to tuning the arrival distribution.

\begin{figure}[h]
    \centering
    \includegraphics[width=0.98\textwidth]{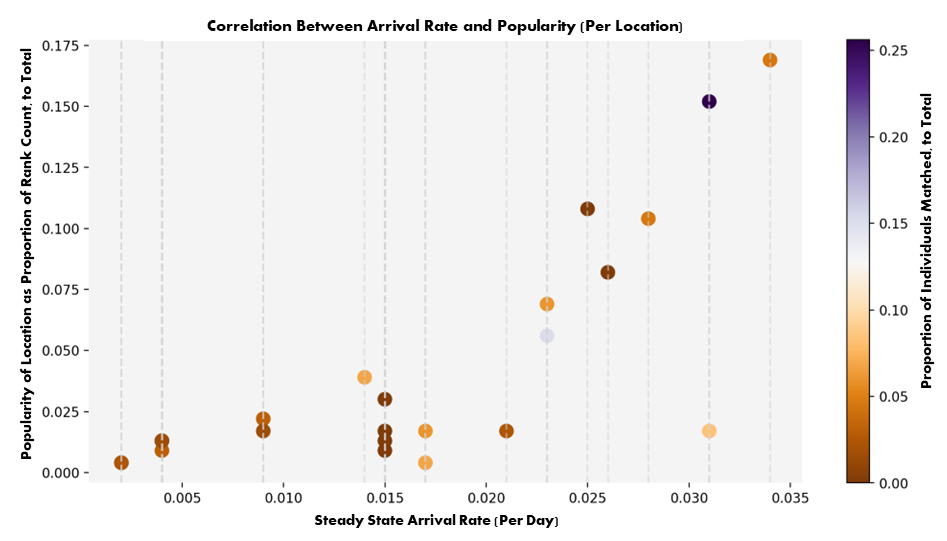}
    \caption{Estimated arrival rates in steady state, for each location validating with empirical popularity and matched individuals}
    \label{pic:SimulationValidation}
\end{figure}

From Figure~\ref{pic:SimulationValidation} we can see that the steady-state arrival rate of the tuned distribution correlates with the empirical probability of locations as measured by their top-three ranking probability, which in turn leads to a larger proportion of the total matches coming from popular locations.

\end{document}